\documentclass{JFM-FLM_Au}

\makeatletter
\def\@j@urnal{}
\makeatother

\renewcommand{\pagelimitfooter}{%
    \hbox to \textwidth{%
        {\cppagefont\thepage}\hfill
    }%
}

\renewcommand{\oddabsfooterflag}{%
    \hbox to \textwidth{%
        \hfill{\cppagefont\thepage}%
    }%
}

\renewcommand{\evenabsfooterflag}{%
    \hbox to \textwidth{%
        {\cppagefont\thepage}\hfill
    }%
}

\makeatother

\usepackage{booktabs}
\usepackage{multirow}
\usepackage{array}
\usepackage{ragged2e}

\usepackage{tabularx}
\usepackage{array}
\usepackage{multirow}

\usepackage{style_ftc}

\lefttitle{S. J.-B. Escudier, F. Temprano-Coleto and H. A. Stone}
\righttitle{Journal of Fluid Mechanics}

\title{Self-similarity of axisymmetric Marangoni flows on a deep viscous subphase}

\author{Sacha J.-B. Escudier\aff{1}, Fernando Temprano-Coleto\aff{1,}\aff{2} \and Howard A. Stone\aff{1}}

\affiliation{\aff{1}Department of Mechanical and Aerospace Engineering, Princeton University, Princeton, NJ 08544, USA
\aff{2}Andlinger Center for the Energy and the Environment, Princeton University, Princeton, NJ 08544, USA}

\corresau{Howard A. Stone, hastone@princeton.edu}

\begin{document}
\maketitle

\begin{abstract}
Local gradients of surface tension caused by inhomogeneous distributions of surfactant molecules are known to generate Marangoni flows, the mechanics of which have important consequences in various areas of science and engineering. In this paper, we investigate the deposition of insoluble surfactants on the surface of a deep, viscous subphase, which propagate axisymmetrically without diffusing. We study these flows from the perspective of self-similarity, i.e., seeking solutions which show universal features on appropriate length and time scales. We first present a closed form, self-similar solution of the first kind for the axisymmetric spreading of surfactants, which is globally valid at long times. We then numerically study `filling' flows of surfactant, where the surfactant distribution is locally depleted such that inward flows are generated to `fill' the surfactant-free area. These flows have the special feature of forming cusp-like structure in finite time, which have been studied extensively in the literature of self-similarity of the second kind. We show that these flows admit a countably infinite number of stable similarity solutions based on the local behavior of the initial surfactant concentration profile. We remark on a few unique features of these results in contrast to filling flows in other physical limits.
\end{abstract}



\par
\vskip14.5pt
\hrule
\vskip18pt


\section{Introduction}\label{sec:Intro}
Surfactant transport remains an important problem throughout various areas of science and engineering, with applications to the environment \citep{fay1969spread, hoult_oil_1972, wokosin_emerging_2022}, biology \citep{gaver1992droplet, angelini_bacillus_2009}, and other technological processes such as microfluidics \citep{stone_engineering_2004} and inkjet printing \citep{de_jong_marangoni_2007}. Localized concentrations of surface-active substances result in gradients of surface tension and, correspondingly, generate shear stresses that induce flow via the Marangoni effect \citep{de_gennes_capillarity_2004}. The continuous redistribution of surfactant molecules along the interface introduces strong coupling between hydrodynamic features of the flow and surfactant transport dynamics, thereby representing a rich class of nonlinear convection problems. 

We consider the spreading of a surfactant monolayer on the infinite surface of a viscous fluid. The mechanics of surfactant replacement therapy first motivated quantitative descriptions of such physical systems, where the fluid subphase is thin relative to the laterally-spreading surfactant front \citep{borgas_monolayer_1988, jensen_insoluble_1992}. These studies showed that the resulting motion is \textit{self-similar}, as the governing equations admit solutions that exhibit universal behavior on appropriate spatial and temporal scales \citep{barenblatt1996scaling}. Likewise, the case of surfactant spreading in other hydrodynamic regimes has been considered \citep{hoult_oil_1972, jensen_spreading_1995, thess1995viscous, crowdy2021viscous}, consistently displaying power-law dependence of the spreading front of surfactant on time (denoted $\ell(t)$ in table \ref{tab:surfactant-front-scaling-table}), a feature consistent with self-similarity. Similarly, `filling' flows generated by surfactant distributions that collapse to fill a region of higher surface tension, have been investigated previously in the literature in various hydrodynamic regimes \citep{jensen_self-similar_1994, thess_two-dimensional_1997, crowdy2021viscous, bickel2022exact, temprano2024self}, and have also been shown to be self-similar. 
\begin{table}
\centering
\renewcommand{\arraystretch}{1.10}

\begin{tabular}{
@{}
>{\raggedright\arraybackslash}p{0.30\linewidth}
>{\raggedright\arraybackslash}p{0.20\linewidth}
>{\raggedright\arraybackslash}p{0.16\linewidth}
@{}
}

\textbf{Geometry}
& \textbf{Constraint}
& \textbf{Scaling of $\ell(t)$} \\[2pt]

\cline{1-3}
\noalign{\vskip 5pt}

\multicolumn{3}{@{}l@{}}{\textbf{Unbounded Stokes flow}} \\
\multicolumn{3}{@{}l@{}}{\textbf{\citep{crowdy2021viscous, bickel2022exact, temprano2024self}}} \\
\addlinespace[5pt]

Planar `strip'
& Constant $M$
& $t^{1/2}$ \\

Planar `front'
& Fixed $\Gamma(r=0)$
& $t$ \\

\addlinespace[3pt]

Axisymmetric `drop'
& Constant $M$
& $t^{1/3}$ \\

Axisymmetric `front'
& Fixed $\Gamma(r=0)$
& $t$ \\

\addlinespace[7pt]

\multicolumn{3}{@{}l@{}}{\textbf{Unbounded inertial flow}} \\
\multicolumn{3}{@{}l@{}}{\textbf{\citep{hoult_oil_1972, foda_spreading_1980, jensen_spreading_1995, dussaud_dynamics_1998}}} \\
\addlinespace[5pt]

Planar `strip'
& Constant $M$
& $t^{1/2}$ \\

Planar `front'
& Fixed $\Gamma(r=0)$
& $t^{3/4}$ \\

\addlinespace[3pt]

Axisymmetric `drop'
& Constant $M$
& $t^{3/8}$ \\

Axisymmetric `front'
& Fixed $\Gamma(r=0)$
& $t^{3/4}$ \\

\addlinespace[7pt]

\multicolumn{3}{@{}l@{}}{\textbf{Thin film flow}} \\
\multicolumn{3}{@{}l@{}}{\textbf{\citep{ahmad_simple_1972, borgas_monolayer_1988, jensen_insoluble_1992}}} \\
\addlinespace[5pt]

Planar `strip'
& Constant $M$
& $t^{1/3}$ \\

Planar `front'
& Fixed $\Gamma(r=0)$
& $t^{1/2}$ \\

\addlinespace[3pt]

Axisymmetric `drop'
& Constant $M$
& $t^{1/4}$ \\

Axisymmetric `front'
& Fixed $\Gamma(r=0)$
& $t^{1/2}$ \\

\end{tabular}

\caption{Summary of known spreading scalings of insoluble surfactant fronts, in the advection-dominated limit (i.e., $\Pen_s \gg 1$). The constraint `Constant $M$' denotes conservation of the total surfactant mass on the interface, and `Constant $\Gamma(r=0)$' denotes a fixed surfactant concentration at the center of the interface. Here, $\ell(t)$ denotes the position of the surfactant front over time, with each entry in the rightmost column showing the power-law time dependence $\ell(t) \propto t^n$, where $n$ is the power-law characteristic of the dynamics. The `strip' (`drop') nomenclature denotes spreading of a fixed amount of surfactant in a planar (axisymmetric) geometry, and `front' denotes spreading from a fixed source of surfactant, as summarized by \citet{berg_marangoni-driven_2009}. If we denote the depth of the fluid subphase $H$, a `thin film flow' indicates $\ell(t) / H \gg 1$, whereas an `unbounded flow' indicates that $\ell(t) / H \ll 1$.}
\label{tab:surfactant-front-scaling-table}
\end{table}

In this paper, we study Marangoni flows on a deep subphase for low-Reynolds-number flow conditions, for axisymmetric surfactant distributions that propagate along a two-dimensional, flat interface. The particularity of this hydrodynamic configuration is that, as first shown by \citet{thess1995viscous}, the velocity field at any given point depends on the entire distribution of surfactant along the domain, and is hence non-local. Despite the complexity this non-locality introduces, recent developments for this physical problem for `strips' of surfactant, i.e., when the flow in the bulk fluid is two-dimensional and interfacial dynamics are one-dimensional (figure \ref{fig:planar_and_axisymmetric_flows}a, c), have identified explicit solutions \citep{crowdy2021viscous, bickel2022exact}, including asymptotic self-similar solutions \citep{temprano2024self}. However, solutions for the axisymmetric problem remain scarce, despite the prevalence surfactant `drops' rather than `strips' in many physical systems (figure \ref{fig:planar_and_axisymmetric_flows}b, d). Here, we study this physical system in the limit of negligible surfactant diffusivity, i.e., a large surface P\'eclet number, $\Pen_s \gg 1$ defined below, as well as surfactant insolubility. We consider the surface velocity field, the surfactant concentration field, and highlight self-similar characteristics of their corresponding solutions.

In section \ref{sec:problem-set-up}, the general problem is presented and the governing equations are given (\S \ref{subsec:gov-eqns}), and a general self-similar structure is posed (\S \ref{subsec:self-sim-structure}). In section \ref{sec:spreading-sol}, a self-similar solution for Marangoni spreading is identified in direct analogy to the work of \citet{bickel2022exact}. In section \ref{sec:filling-sol}, filling flows are considered and analytical features unique to these solutions are identified (\S \ref{subsec:analytical-strat}). Subsequently, we proceed to study these flows numerically (\S \ref{subsec:num-analysis-methods}), and show that a spectrum of self-similar solutions exists depending on the family of initial conditions (\S \ref{subsec:num-analysis-results}). We discuss these numerical results (\S \ref{subsec:discussion-filling}), and summarize all findings in section \ref{sec:conclusion}.

\section{Problem set-up} \label{sec:problem-set-up}
\subsection{Governing equations} \label{subsec:gov-eqns}
\begin{figure}
  \centerline{\includegraphics[width=1.0\textwidth]{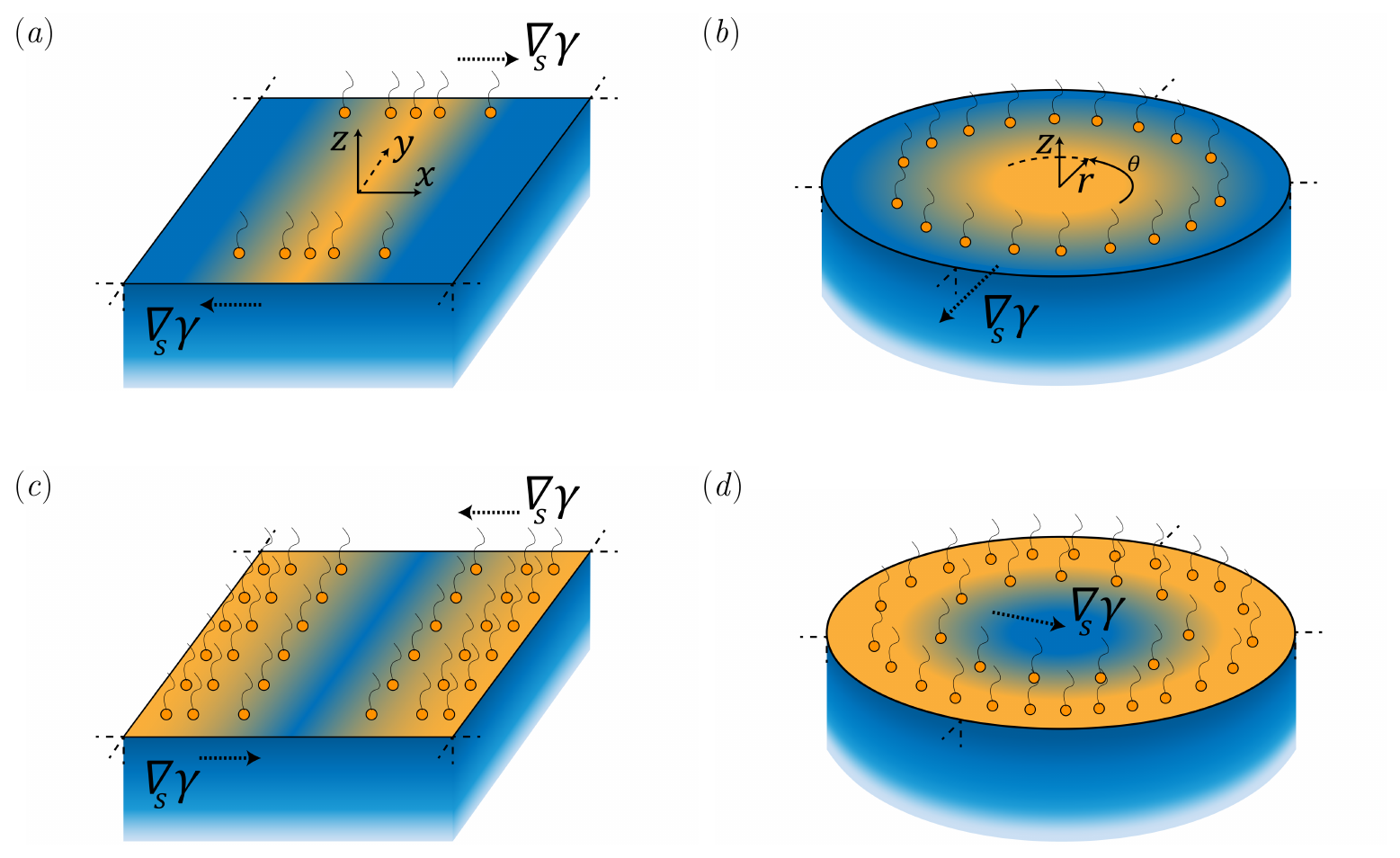}}
  \caption{Comparison of Marangoni flows on a deep fluid subphase in the low-Reynolds-number limit. (\textit{a}) Spreading of a planar `strip' of surfactant \citep{thess1995viscous, crowdy2021viscous, bickel2022exact, temprano2024self}; (\textit{b}) spreading of an axisymmetric `drop' of surfactant \citep[][this work]{bickel2022exact}; (\textit{c}) filling  of a planar initial distribution of surfactant \citep{thess1995viscous, crowdy2021viscous, bickel2022exact, temprano2024self}; (\textit{d}) filling of an axisymmetric initial distribution of surfactant (this work).}
  \label{fig:planar_and_axisymmetric_flows}
\end{figure}

We consider Marangoni flows initiated by an insoluble surfactant on the upper free surface of an infinitely deep, Newtonian fluid subphase (figure \ref{fig:planar_and_axisymmetric_flows}). We let
$\ell_c$ denote a characteristic length scale of the surfactant concentration field, $u_c$ a characteristic speed set by Marangoni convection, $\gamma_0$ the nominal surface tension of the interface, $D_s$ the surface diffusivity of the surfactant, $\rho$ the density of the subphase, and $\mu$ its dynamic viscosity. We first note the magnitudes of the Reynolds, capillary, and surface P\'eclet numbers,
\begin{equation}
    \Rey = \frac{\rho u_c \ell_c}{\mu} \ll 1, \quad Ca = \frac{\mu u_c}{\gamma_0} \ll 1, \quad \Pen_s = \frac{u_c\ell_c}{D_s} \gg 1,
\end{equation}
which reflect that the flow is viscously dominated, interface deflection is negligible, and surfactant advection is dominant, respectively. For this physical problem, it is convenient to use cylindrical coordinates, so the equations of motion for the flow field $\boldsymbol{u} = \boldsymbol{u}(r, z, t)$, pressure field $p=p(r,z,t)$, and surfactant concentration field $\Gamma=\Gamma(r,t)$ are
\begin{subequations}
    \begin{align}
        \bnabla p = \mu \nabla^2 \boldsymbol{u}, \quad \bnabla \bcdot \boldsymbol{u} = 0, \label{eq:stokes+contd} \\
        \frac{\p \Gamma}{\p t} + \bnabla_s \bcdot (\left.\boldsymbol{u}\right|_{z=0}\Gamma) = 0, \label{eq:surfactant-transport}
    \end{align}
\end{subequations}
where $\bnabla_s$ represents the surface gradient operator along the interface. For convenience, we subsequently refer to the surface velocity as $u_s = \Vert \left( \boldsymbol{I} - \boldsymbol{n}\boldsymbol{n}\right) \bcdot \left.\boldsymbol{u}\right|_{z=0} \Vert$, where $\boldsymbol{n}$ is the surface normal vector \citep{stone_simple_1990} and $z=0$ denotes the upper free surface, see figure \ref{fig:planar_and_axisymmetric_flows}. The coupling of the hydrodynamics to the surfactant transport problem is achieved through a boundary condition of stress at the interface and an equation of state, namely
\begin{subequations}
    \begin{align}
        \left.\boldsymbol{e}_r \bcdot \mathsfbi{T} \bcdot \boldsymbol{e}_z\right|_{z=0} =\boldsymbol{e}_r \bcdot \bnabla_s\gamma, \label{eq:stress-boundary-condition}\\
        \gamma(\Gamma) = \gamma_0 - a\Gamma, \label{eq:eq-of-state}
    \end{align}
\end{subequations}
where $\mathsfbi{T}$ is the stress tensor associated with the subphase, $\boldsymbol{e}_i$ denotes a unit vector, with, for a planar surface, $\boldsymbol{e}_z=\boldsymbol{n}$, and, for an assumed linear variation of surface tension with surface concentration, $a = |\dd\gamma/\dd\Gamma|$ is a constant. We note that equation \eqref{eq:eq-of-state} is consistent with dilute concentrations of surfactant \citep{manikantan2020surfactant}. We proceed to non-dimensionalize the problem according to
\begin{equation}
    r \mapsto \ell_c r, \quad \Gamma \mapsto \Gamma_c \Gamma, \quad u \mapsto \left(\frac{a\Gamma_c}{2\mu}\right)u, \quad t \mapsto \left(\frac{2\mu \ell_c}{a\Gamma_c}\right)t,
\end{equation}
where $\Gamma_c$ is a characteristic surfactant concentration and is typically selected as the maximum value of the initial concentration distribution $\Gamma_0(r) = \Gamma(r,t=0)$, i.e., $\Gamma_c := \max(\Gamma_0)$. In finding solutions, we focus on equation \eqref{eq:surfactant-transport} only by supplementing it with a `closure' relation for $u_s(\Gamma)$, which is derived from the Stokes and continuity equations \eqref{eq:stokes+contd}, as well as the stress boundary condition \eqref{eq:stress-boundary-condition} and the equation of state \eqref{eq:eq-of-state}. In particular, \citet{thess1995viscous} first showed that a transformation of the governing equation for the velocity leads to a non-local closure law between the surface velocity and the surfactant concentration field, see a detailed derivation by \citet{thess_two-dimensional_1997}. For Marangoni flows in planar geometries (figure \ref{fig:planar_and_axisymmetric_flows}a, c), the closure relation reduces to a Hilbert transform \citep{thess1995viscous}, for which \citet{crowdy2021viscous} has shown that the introduction of an upper analytic function $\psi = u_s + \mathrm{i}\Gamma = H[\Gamma] + \mathrm{i}\Gamma$ transforms \eqref{eq:surfactant-transport} into an inviscid Burgers equation,
\begin{equation}
    \frac{\p \psi}{\p t} + \psi \frac{\p \psi}{\p x} = 0, \label{eq:Burgers-equation}
\end{equation}
where $H[\Gamma] = \dashint{-\infty}{\infty} \frac{\Gamma(x', t)}{x-x'} \dd x'$, with the dash indicating the integral should be interpreted in the principal value sense. The unique feature of the transport equation \eqref{eq:Burgers-equation} is its locality, meaning that the introduction of a complex-valued function $\psi$ effectively transforms the original problem from an integro-differential equation to a point-wise partial differential equation. Subsequently, solutions to \eqref{eq:Burgers-equation} can be found via the method of characteristics \citep{crowdy2021viscous, bickel2022exact} or through asymptotic self-similarity \citep{temprano2024self}. In contrast, as shown by \citet{bickel2022exact}, the closure relation for surfactant transport along an axisymmetric interface reduces to
\begin{equation}
    u_s(r,t) = \mathcal{H}_1^{-1} \left[ \mathcal{H}_0 \left[ \Gamma(r,t) \right]\right] = \int_0^\infty s J_1(s r) \left(\int_0^\infty r' J_0(s r')\Gamma(r',t) \dd{r'}\right)\dd{s}, \label{eq:closure-axisymmetric}
\end{equation}
where $\mathcal{H}_1^{-1}$ and $\mathcal{H}_0$ are the inverse first-order and zeroth-order Hankel transforms, respectively. Unlike the introduction of the Hilbert transform in the planar flow problem, it is not known whether locality of \eqref{eq:surfactant-transport} can be recovered using the closure relation \eqref{eq:closure-axisymmetric} in the axisymmetric problem. We note that \eqref{eq:closure-axisymmetric} can be expressed using a single integral as shown in appendix \ref{apx:numerics_closure}, the form of which is particularly useful in the numerical evaluation of the closure relation. Generally, we resort to analyzing the full set of equations
\begin{subequations}
    \begin{align}
        0 &= \frac{\p \Gamma}{\p t} + \frac{1}{r} \frac{\p}{\p r} \left( ru_s\Gamma\right), \\
        \quad u_s(r,t) &= \int_0^\infty  s J_1(s r) \left( \int_0^\infty  r' J_0(s r')\Gamma(r',t)  \dd{r'}\right)  \dd{s},  \label{eq:pde-closure-only}
    \end{align} \label{eq:pde-with-closure}
\end{subequations}
and seek self-similar solutions for $u_s(r,t)$ and $\Gamma(r,t)$. 

\subsection{Self-similar structure} \label{subsec:self-sim-structure}
In an effort to develop a self-similar framework, we pose an ansatz of the form
\begin{subequations}
\begin{align}
        \Gamma(r,t) &= A|t-t_*|^\alpha C(\eta), \\
        u_s(r,t) &= A|t-t_*|^\alpha U(\eta), \\
        \eta &= \frac{r}{B|t-t_*|^\beta},
    \end{align} \label{eq:similarity-ansatz}
\end{subequations}
where $t_*$ is a reference time, $A$ and $B$ are constants, and $C(\eta)$ and $U(\eta)$ are the similarity solutions for the concentration and velocity, respectively. Upon substitution of \eqref{eq:similarity-ansatz} into \eqref{eq:pde-with-closure}, we obtain the system of equations
\begin{subequations}
    \begin{align}
        0 &= (\beta - 1)C - \beta \eta \frac{\dd C}{\dd\eta} + \sgn{t-t_*}\frac{A}{B\eta} \frac{\dd}{\dd\eta} \left( \eta U C\right), \label{eq:similarity-ODE-a}\\
        \quad U(\eta) &= \int_0^\infty  s J_1(s\eta) \left( \int_0^\infty  \eta' J_0(s\eta')C(\eta')  \dd{\eta'}\right)  \dd{s}. \label{eq:closure-similarity}
    \end{align} \label{eq:similarity-ODE}
\end{subequations}
Given the ansatz \eqref{eq:similarity-ansatz} and the form of the governing equations, the exponents $\alpha$ and $\beta$ follow the constraint $\alpha = \beta - 1$ in order for the explicit time dependence of \eqref{eq:similarity-ODE} to vanish \citep{eggers2015singularities}. It remains to find the value of the similarity exponent $\beta$, which is determined by the physical features of the system at hand. It is often possible to uniquely determine $\beta$ by supplementing the problem with a known constraint (such as the conservation of a finite surfactant mass), which is the case of \textit{self-similarity of the first kind}. It is, however, also possible that the value of $\beta$ cannot be determined from dimensional considerations, in which case \eqref{eq:similarity-ODE} becomes a nonlinear eigenvalue problem, corresponding to \textit{self-similarity of the second kind} \citep{barenblatt1996scaling, Eggers_2009}. In the context of axisymmetric viscous Marangoni flows, we discuss a case of self-similarity of the first kind in section \ref{sec:spreading-sol}, and a case of the second kind in section \ref{sec:filling-sol}.

\section{Axisymmetric Marangoni spreading solution}\label{sec:spreading-sol}
\begin{figure}
  \centerline{\includegraphics[width=1.0\textwidth]{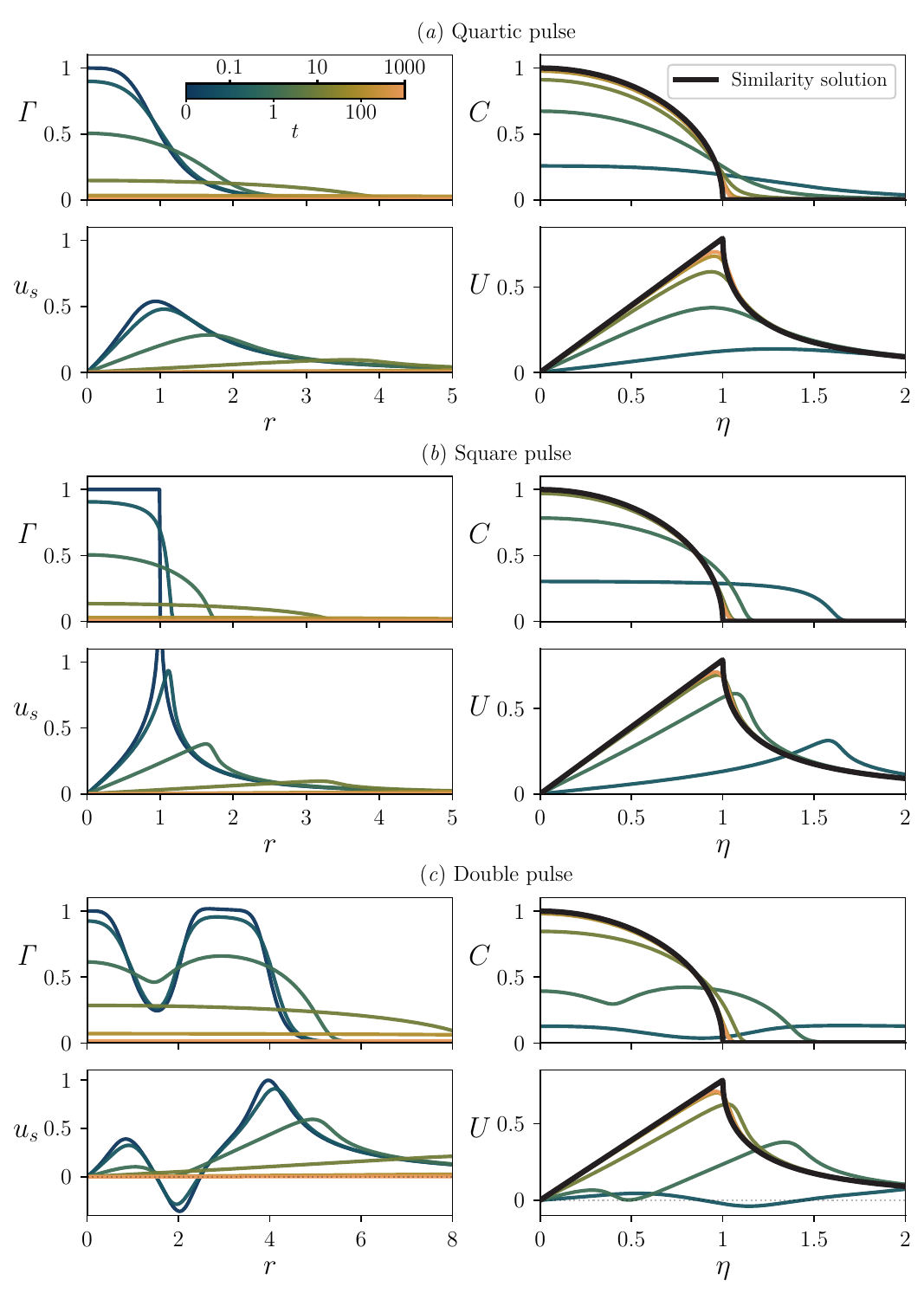}}
  \caption{Different spreading pulses of surfactant converge to the same similarity solution for axisymmetric Marangoni flows. The left column shows the numerical solution of surfactant concentration and surface velocity profiles, and the right column re-scales these solutions in similarity space to show the corresponding collapse onto the similarity solution \eqref{eq:spreading-full-similarity-solution}. The different initial concentration profiles are (\textit{a}) $\Gamma_0 = 1/(1+r^4)$, (\textit{b}) $\Gamma_0 = H(1-r)$ with $H(\, \cdot \,)$ the Heaviside function, and (\textit{c}) $\Gamma_0 = 1/(1+r^4) + 1/(1 + (r-3)^6)$.}
  \label{fig:spreading-similarity-collapse}
\end{figure}
We first consider the case of an axisymmetric, spreading Marangoni flow (figure \ref{fig:planar_and_axisymmetric_flows}b), for which a solution was previously identified by \citet{bickel2022exact} through an ansatz informed by known similarity solutions of the nonlinear porous media equations \citep{caffarelli_nonlinear_2011, biler2011barenblatt}. In this section, we derive this solution through self-similar analysis, making use of the non-local closure relation \eqref{eq:closure-axisymmetric} to form a solvable system of dual integral equations. Since we expect the solution to be asymptotically valid for $t \gg 1$, we consider $t_* = 0$ in equation \eqref{eq:similarity-ODE} for spreading flows.

To begin, we note that the mass of the surfactant is conserved on the interface at all times, so we have the integral constraint
\begin{equation}
    \int_0^\infty r\Gamma \dd{r} = M = \text{const}., \label{eq:integral-constraint}
\end{equation}
where $M$ represents the mass of surfactant (up to a factor of $2\upi$). Substitution of the self-similar ansatz \eqref{eq:similarity-ansatz} into equation \eqref{eq:integral-constraint} gives
\begin{equation}
    t^{3\beta - 1}\int_0^\infty  \eta C \dd{\eta} = \frac{M}{A B^2} \ \Rightarrow \ \beta = \frac{1}{3}.
\end{equation}
For this value of $\beta$, \eqref{eq:similarity-ODE-a} can be integrated exactly, yielding
\begin{equation}
    C(\eta)\left(U(\eta) -\frac{B \eta}{3A} \right) = 0, \label{eq:spreading-ODE-condition}
\end{equation}
from which it is clear that, to satisfy the governing equations, either the surfactant concentration must be zero along the interface, or the surface velocity must scale linearly with the radius. Since we only consider solutions for $U(\eta)$ that are bounded as $\eta \to{\infty}$ to be physically relevant, we conclude from \eqref{eq:spreading-ODE-condition} that the solution must be compactly supported past a critical value $\eta_c$, and identify the piece-wise features of the full solution as
\begin{equation}
C(\eta)=
\begin{cases}
  C_-, & \eta \leq \eta_c,\\
  0, & \eta > \eta_c,
\end{cases}
\qquad
U(\eta)=
\begin{cases}
  B\eta/\left(3A\right), & \eta \leq \eta_c,\\
  U_+, & \eta > \eta_c.
\end{cases}  \label{eq:spreading-partial-similarity-solution}
\end{equation}
where $C_-(\eta)$ and $U_+(\eta)$ are the unknown components of the piece-wise solutions for $C(\eta)$ and $U(\eta)$, respectively. We now remark that the closure relation may be rewritten as
\begin{equation}
    U = -\frac{\dd}{\dd\eta} \mathcal{H}_0^{-1} \left[ s^{-1} \mathcal{H}_0 \left[ C \right]\right] \ \Rightarrow \ \mathcal{H}_0\left[\Tilde{U}\right] = s^{-2} \phi(s), \quad \mathcal{H}_0\left[C\right] = s^{-1}\phi(s), \label{eq:rewritten-closure}
\end{equation}
for some auxiliary function $\phi(s)$. For convenience, we also defined $\Tilde{U} = K - \int^\eta U(\eta') \dd{\eta'}$ where $K$ is some constant. Equivalently, \eqref{eq:rewritten-closure} may be explicitly written as 
\begin{subequations}
    \begin{align}
        C(\eta) &= \int_0^\infty J_0(s\eta) \phi(s) \dd{s}, \\
        \Tilde{U}(\eta) &= \int_0^\infty s^{-1} J_0(s\eta) \phi(s) \dd{s},
    \end{align}\label{eq:dual-integral-equations}
\end{subequations}
where the system of dual integral equations \eqref{eq:dual-integral-equations} has a known solution provided by \citet{sneddon1966mixed}, given the known piece-wise features of the full solution provided by \eqref{eq:spreading-partial-similarity-solution}. Combined with the requirement of continuity of the flow profile $U(\eta)$, this leads to the solution
\begin{equation}
C(\eta)=
\begin{cases}
  \sqrt{1-\eta^2}, & \eta \leq 1,\\
  0, & \eta > 1,
\end{cases}
\qquad
U(\eta)=
\begin{cases}
  \frac{\upi}{4}\eta, & \eta \leq 1,\\
    \frac{1}{2\eta} \left[\eta^2 \csc^{-1}\eta - \sqrt{\eta^2-1}\right], & \eta > 1,
\end{cases} \label{eq:spreading-full-similarity-solution}
\end{equation}
where the constants appearing in \eqref{eq:similarity-ansatz} have been fixed to $A = (16M / (3 \upi^2))^{1/3}$ and $B = (9\upi M/4)^{1/3}$, to obtain simple prefactors. Figure \ref{fig:spreading-similarity-collapse} shows, for several initial conditions, that the numerical solutions to \eqref{eq:pde-with-closure} collapse onto the self-similar solutions \eqref{eq:spreading-full-similarity-solution} for long times $t \gg 1$. The numerical procedure used to solve \eqref{eq:pde-with-closure} is identical to the one used in the analysis of `filling' flows, and is outlined in section \ref{sec:filling-sol}.

\section{Filling solutions} \label{sec:filling-sol}
\subsection{Analytical features} \label{subsec:analytical-strat}
\begin{figure}
  \centerline{\includegraphics[width=1.0\textwidth]{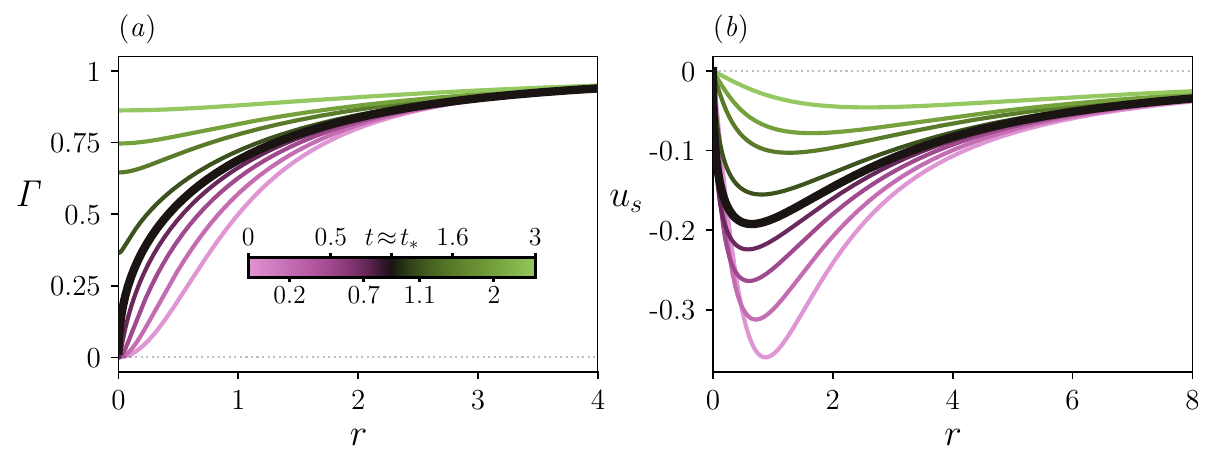}}
  \caption{Numerical solution of axisymmetric filling flows. (\textit{a}) Time evolution of the surfactant concentration profile $\Gamma(r,t)$ and (\textit{b}) of the surface velocity field $u_s(r,t)$, subject to the initial condition $\Gamma_0 = r^2 / (1+r^2)$. The purple curves denote `closure' (pre-singularity) profiles, and the green curves denote `leveling' (post-singularity) profiles. The black curves denote the profiles of $\Gamma$ and $u_s$ at $t \approx t_*$.}
  \label{fig:filling-example}
\end{figure}
Consider now the axisymmetric closure of a surfactant-free area in the midst of a contaminated surface, which we previously referred to as filling flows (figure \ref{fig:planar_and_axisymmetric_flows}d). Here, we restrict ourselves to cases where $\Gamma_0(r=0) = 0$, as considered by previous studies \citep{thess1995viscous, crowdy2021viscous, bickel2022exact, temprano2024self}. In this case, the concentration profile has a fixed value in the far-field, such that the mass of surfactant along the interface is not conserved over time. The similarity exponent then cannot be specified by dimensional analysis, and its value is instead obtained by solving a nonlinear eigenvalue problem consisting of the governing similarity equations \eqref{eq:similarity-ODE} and boundary conditions, a notion known as self-similarity of the second kind \citep{barenblatt1996scaling, Eggers_2009}. We proceed to solve the problem given by \eqref{eq:pde-with-closure} numerically to identify whether self-similar characteristics exist for axisymmetric filling flows. If self-similar solutions exist, we can expect to recover the following characteristics across different numerical simulations:
\begin{enumerate}
    \item \hspace{1mm} By direct analogy with Marangoni filling flows in other regimes, we expect that both $\Gamma(r,t)$ and $u_s(r,t)$ will develop a singularity in their first derivative at $r=0$ at some reference time $t=t_*$; see figure \ref{fig:filling-example} for typical singular profiles of $\Gamma(r,t)$ and $u_s(r,t)$ in an axisymmetric geometry. This reference time $t_*$ is unique to each initial condition, and is not a universal feature of the problem.
    \item \hspace{1mm} We expect that both $\Gamma(r,t)$ and $u_s(r,t)$ will have a `closing' (pre-singularity) similarity solution as they approach their singular form, and a `leveling' (post-singularity) similarity solution as they relax to their far-field value, both valid in the limit of $|t-t_*|\ll 1$ according to the ansatz in \eqref{eq:similarity-ansatz}. We also note that, owing to the functional invariance of the Hankel transforms in \eqref{eq:closure-similarity} to power laws (i.e., $\mathcal{H}_1^{-1} \mathcal{H}_0 \left[\eta^\theta \right] \propto \eta^\theta$ for a constant $\theta$) \citep{NIST:DLMF}, the similarity profiles $C(\eta)$ and $U(\eta)$ will share the same similarity exponent. 
\end{enumerate} 
We proceed to measure the value of the similarity exponent $\beta$ subject to different initial concentration profiles by (1)~analyzing the pre-singular behavior of $\Gamma(r,t)$ and $u_s(r,t)$ as they become singular and (2)~analyzing the post-singular behavior of $\Gamma(r,t)$ as it relaxes to its far-field value.

\begin{figure}
  \centerline{\includegraphics[width=1.0\textwidth]{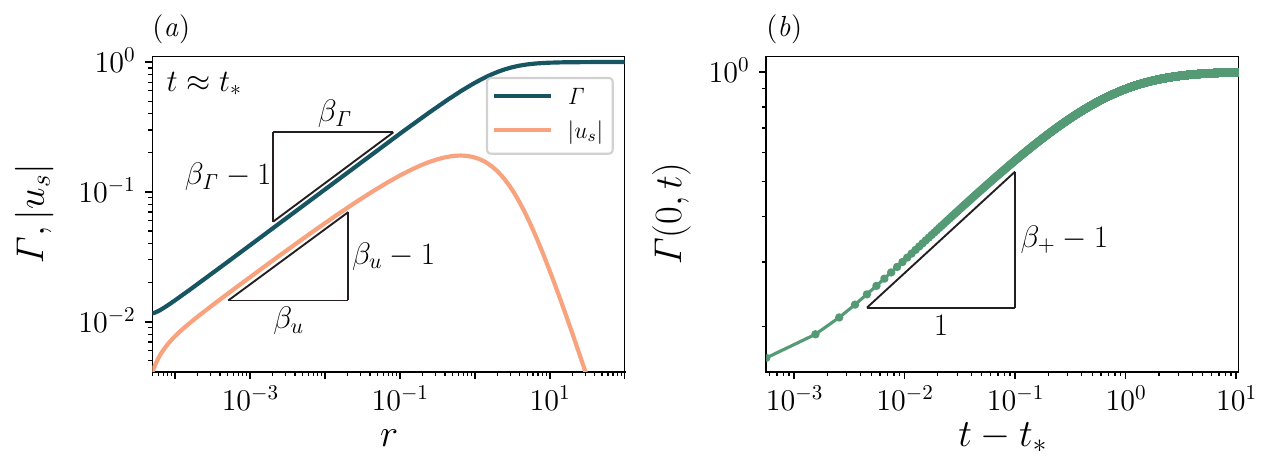}}
  \caption{Numerically obtained concentration and surface velocity profiles from which the value of $\beta$ is extracted, using the relations established in \eqref{eq:gamma-us-propto-r} and \eqref{eq:gamma-propto-t}. (\textit{a}) Power-law profiles for $\Gamma(r,t)$ and $u_s(r,t)$ at $t\approx t_*$, for the initial condition $\Gamma_0 = r^2 / (1+r^2)$. (\textit{b}) Power-law profiles of the leveling origin $\Gamma(r=0,t)$ versus $t - t_*$, for a second initial condition $\Gamma_0 = H(r - 1)$.}
  \label{fig:selection-criterion}
\end{figure}

(1)~For $C(\eta)$ and $U(\eta)$, we expect a known far-field condition for both closing and leveling similarity solutions, stemming from the requirement that they must match to the same non-self-similar outer region \citep{Eggers_2009, eggers2015singularities}. This is expressed as 
\begin{subequations}
    \begin{align}
         C(\eta) &\sim k_c \eta^\frac{\beta - 1}{\beta} \text{ as } \eta \to{\infty}, \\
         U(\eta) &\sim k_u \eta^\frac{\beta - 1}{\beta} \text{ as } \eta \to{\infty},
    \end{align}\label{eq:boundary-conditions-filling}
\end{subequations}
where $k_c$ and $k_u$ are constants. We may take advantage of \eqref{eq:boundary-conditions-filling} by substituting it into the self-similar ansatz \eqref{eq:similarity-ansatz} to express $\Gamma(r,t)$ and $u_s(r,t)$ close to the singular time as
\begin{subequations}
    \begin{align}
         \Gamma(r,t) &\propto r^{\frac{\beta - 1}{\beta}} \text{ at } t \approx t_*,\label{eq:gamma-propto-r} \\
         u_s(r,t) &\propto r^{\frac{\beta - 1}{\beta}} \text{ at } t \approx t_*. \label{eq:us-propto-r}
    \end{align} \label{eq:gamma-us-propto-r}
\end{subequations}
from which the value of $\beta$ can be deduced. Note that we expect the relations established in \eqref{eq:gamma-us-propto-r} to be valid for $|t-t_*|^\beta \ll r \ll 1$, consistent with the asymptotic behavior of the similarity functions established in \eqref{eq:boundary-conditions-filling}. While both curves should theoretically share the same exponent, we note that, in practice, their values will be slightly different due to numerical errors. As a result, we extract $\beta$ from each profile of $\Gamma$ and $u_s$ as $\beta_\Gamma$ and $\beta_u$, respectively (figure \ref{fig:selection-criterion}a), and later refer to the average of their measurements as the `closing' value of $\beta$ (table \ref{tab:pre-and-post-singularity-beta}).

(2)~While the surfactant concentration is zero at the origin prior to the singularity, we expect it to become finite, and eventually level-off to its far-field value, past its singular time (i.e., $\Gamma(r=0, t> t_*) > 0$). The corresponding similarity solution \eqref{eq:similarity-ansatz} evaluated at the origin then follows
\begin{equation}
    \Gamma(0,t) = A|t - t_*|^{\beta - 1} C(0) \ \Rightarrow \ \Gamma(0,t) \propto |t - t_*|^{\beta - 1}, \label{eq:gamma-propto-t}
\end{equation}
from which the value of $\beta$ may again be deduced and is extracted from the evolution of $\Gamma(0,t)$ as $\beta_+$ (figure \ref{fig:selection-criterion}b). We again stress that $\beta_+$ should theoretically be equal to $\beta_\Gamma$ and $\beta_u$, but will be slightly different in practice due to numerical uncertainties. This measurement is listed as the `leveling' value of $\beta$ in table \ref{tab:pre-and-post-singularity-beta}. The relations established in \eqref{eq:gamma-us-propto-r} and \eqref{eq:gamma-propto-t} then provide the necessary numerical criteria to identify the singular time $t_*$ and extract $\beta$ from the simulations, as explained in \S\ref{subsec:num-analysis-methods}.

\subsection{Numerical methods} \label{subsec:num-analysis-methods}
\begin{figure}
  \centerline{\includegraphics[width=1.0\textwidth]{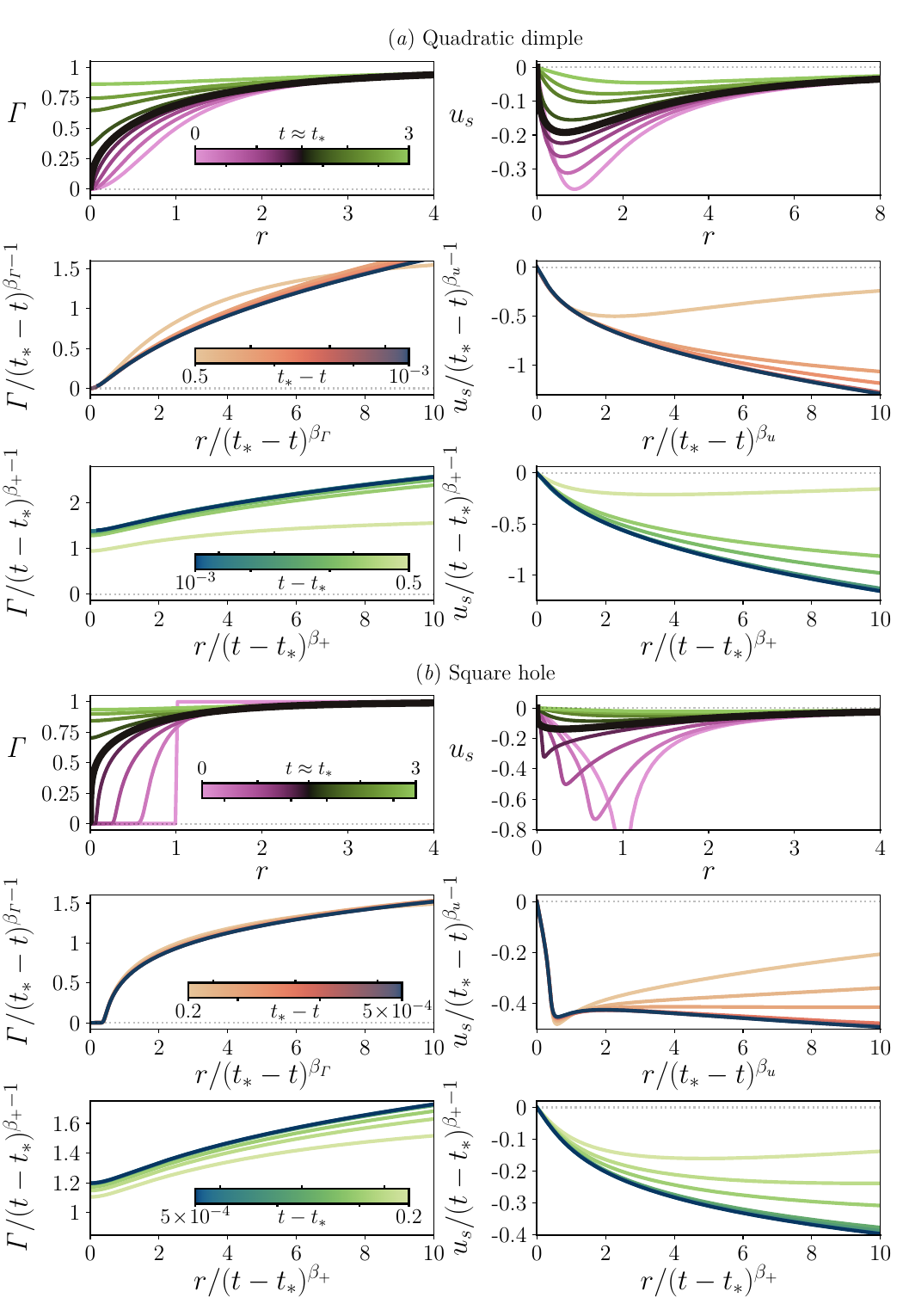}}
  \caption{Different filling pulses and corresponding collapse in similarity space for axisymmetric Marangoni flows. The initial concentration profiles are (\textit{a}) $\Gamma_0 = r^2/(1+r^2)$, and (\textit{b}) $\Gamma_0 = H(r - 1)$. The values $\beta_\Gamma$, $\beta_u$, and $\beta_+$ are extracted from the singular structure associated with each initial condition, as shown in figure \ref{fig:selection-criterion}.}
  \label{fig:filling-similarity-collapse}
\end{figure}
To numerically evaluate the system of equations (\ref{eq:pde-with-closure}), a finite-volume scheme along a one-dimensional grid is implemented. Since we expect the singularity to occur at $r=0$ by symmetry, the cell centers are geometrically spaced at positions $r_n$ such that $r_{n+1} / r_n = \lambda >1 $, allowing for the efficient evaluation of \eqref{eq:closure-axisymmetric} in Fourier space (see appendix \ref{apx:numerics_closure}), but also to resolve the power-law structure of the singularity as finely as possible \citep{burton_two-dimensional_2007, eggers2015singularities}. Across each trial, the grid refinement is varied by directly controlling the position of the first cell center and the number of grid points. Each simulation is explicitly evolved forward in time with a time step consistent with a Courant-Friedricks-Lewy condition much lower than one \citep{hinch2020numerical}. Since we expect both functions to share the same similarity exponent (i.e., theoretically \eqref{eq:gamma-propto-r} and \eqref{eq:us-propto-r} should share the same $\beta$), we also expect to measure the same power-law behavior for $\Gamma(r,t)$ and for $u_s(r,t)$ as $t\to{t_*}$. We therefore estimate $t_*$ by running each simulation forward in time until the logarithmic slopes of $\Gamma(r,t)$ and $u_s(r,t)$ agree most closely over a radial interval $I$ spanning a prescribed length. The selection criterion for $t_*$ and the optimal fitting interval $I_*$ are expressed as
\begin{equation}
    (t_*, I_*) := \underset{(t,I)}{\arg\min}\, \left\langle\left|\frac{\p\log \Gamma}{\p \log r} - \frac{\p \log |u_s|}{\p \log r}\right|\right\rangle_I, \label{eq:pre-singularity-selection-criterion}
\end{equation}
where the $\langle\, \cdot\, \rangle_I$ operator denotes spatial averaging over the radial interval $I$. The exponents $\beta_\Gamma$ and $\beta_u$ are then extracted directly from their respective curves at the singular time $t_*$ over the optimal interval $I_*$ specified by \eqref{eq:pre-singularity-selection-criterion}, as shown in figure \ref{fig:selection-criterion}a. Similarly, the similarity exponent $\beta_+$ is extracted by fitting the post-singular concentration profile at the origin over the best-fit temporal interval, as indicated in figure \ref{fig:selection-criterion}b. Across all numerical trials for all initial conditions, closing and leveling values of $\beta$ were in good agreement, as shown in table \ref{tab:pre-and-post-singularity-beta}. 

\subsection{Numerical results} \label{subsec:num-analysis-results}

\begin{table}
\centering
\renewcommand{\arraystretch}{1.10}

\begin{tabular*}{0.86\linewidth}{
@{\extracolsep{\fill}}
l
c
c
@{}
}

$\boldsymbol{\Gamma_0(r)}$
& \textbf{`Closing' $\beta$}
& \textbf{`Leveling' $\beta$} \\[2pt]

\cline{1-3}
\noalign{\vskip 5pt}

$r^2/(1+r^2)$
& $1.747 \pm 0.018$
& $1.734 \pm 0.034$ \\

$1-e^{-r^2}$
& $1.746 \pm 0.020$
& $1.732 \pm 0.037$ \\

$\tanh^2 r$
& $1.748 \pm 0.022$
& $1.733 \pm 0.034$ \\

\addlinespace[4pt]

$r^4/(1+r^4)$
& $1.302 \pm 0.013$
& $1.301 \pm 0.006$ \\

$1-e^{-r^4}$
& $1.296 \pm 0.010$
& $1.296 \pm 0.009$ \\

$\tanh^4 r$
& $1.315 \pm 0.019$
& $1.317 \pm 0.003$ \\

\end{tabular*}

\caption{Numerically determined values of $\beta$ for the initial profiles shown in figure \ref{fig:varying_behavior_powerlaws}a, 
b, based on the power-law behavior of $\Gamma(r, t)$ and $u_s(r,t)$ leading to the singularity (`closing') and the behavior of $\Gamma(r=0,t)$ beyond the singularity (`leveling'). Note that the above results are grid-independent estimates, which were extrapolated according to the procedure outlined in appendix \ref{apx:numerics_extrapolation}.}

\label{tab:pre-and-post-singularity-beta}
\end{table}
\begin{figure}
  \centerline{\includegraphics[width=1.0\textwidth]{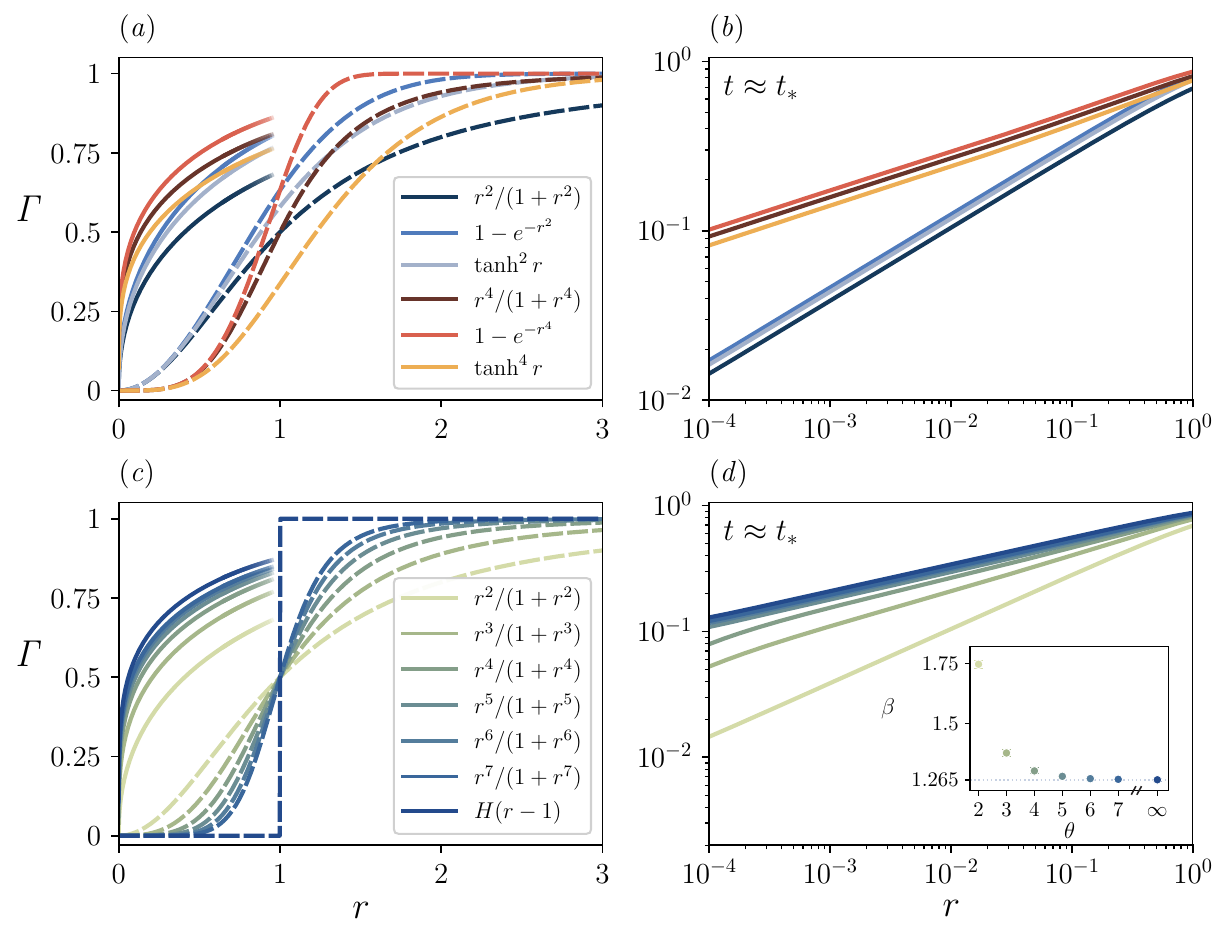}}
  \caption{Initial and singular surfactant concentration profiles along the interface for filling flows. (\textit{a}) Time evolution of different concentration profiles with similar asymptotic behavior near the origin. The dashed lines denote the initial concentration profiles $\Gamma_0(r)$ at $t=0$, and the solid lines denote the numerical solutions at $t\approx t_*$ corresponding to each of the profiles. (\textit{b}) Power-law behavior near the origin for each solution in panel (\textit{a}) at time $t\approx t_*$. (\textit{c}) Time evolution of different concentration profiles with polynomial behavior of increasing order at the origin. The dashed lines denote the concentration profiles $\Gamma_0(r)$ at $t=0$, and the solid lines denote the numerical solutions at $t\approx t_*$. (\textit{d}) Singular profile of each initial condition in panel (\textit{c}) near the origin. Inset shows the dependence of `closing' values of the similarity exponent $\beta$ on the near-origin behavior of $\Gamma_0$, where we consider $\Gamma_0 \sim kr^\theta \text{ as } r \to{0}$.}
  \label{fig:varying_behavior_powerlaws}
\end{figure}
Using the measured values of $\beta_\Gamma, \beta_u, \beta_+$, the numerical solutions to $\Gamma(r,t)$ and $u_s(r,t)$ can be rescaled in similarity space, as shown in figure \ref{fig:filling-similarity-collapse}. Consistent with the self-similar solutions of filling flows in planar geometries \citep{temprano2024self}, concentration and surface velocity profiles collapse onto a power-law structure as $|t-t_*| \to{0}$, the structure of which differs across initial conditions. It is also clear from the results displayed in Figure \ref{fig:filling-similarity-collapse} that the similarity solutions of \eqref{eq:similarity-ODE} admit multiple values of $\beta$, depending on features of the initial condition.    

\begin{figure}
  \centerline{\includegraphics[width=1.0\textwidth]{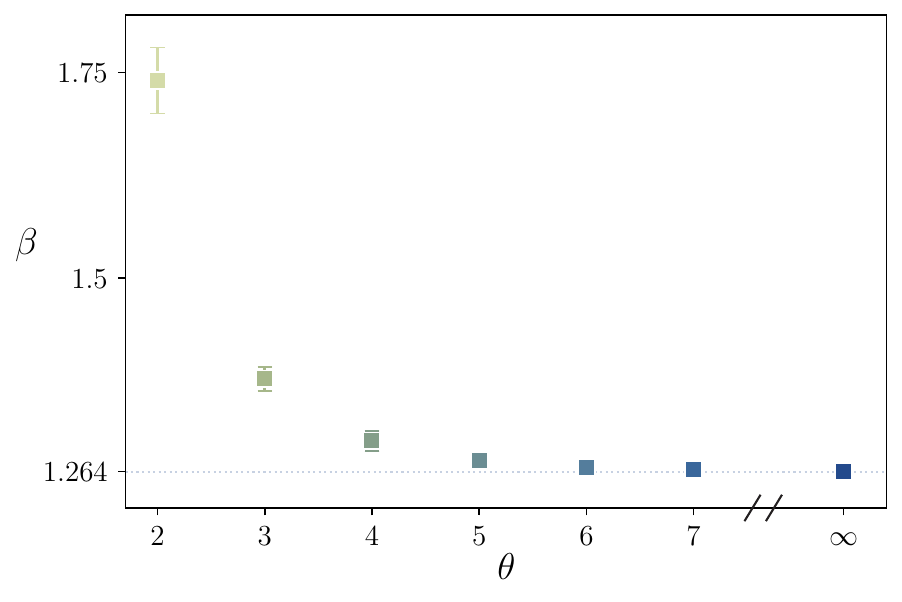}}
  \caption{Dependence of the similarity exponent $\beta$ on the order of the polynomial behavior of the initial condition $\Gamma_0$ at the origin, for both `closing' (figure \ref{fig:selection-criterion}a) and `leveling' (figure \ref{fig:selection-criterion}b) measurements.}
  \label{fig:beta-vs-theta}
\end{figure}

To further investigate the eigenspectrum of \eqref{eq:similarity-ODE}, i.e., the possible values of $\beta$, we computed the time evolution of several initial concentration profiles up to their singular time $t=t_*$, as shown in figure \ref{fig:varying_behavior_powerlaws}. We considered initial conditions with different power-law behaviors at the origin such that $\Gamma\sim{k}r^\theta$ as $r \to{0}$, with $\theta$ an integer and $k$ a constant. We first compared locally quadratic ($\theta=2$) against flatter profiles ($\theta>2$) (figure \ref{fig:varying_behavior_powerlaws}a), and observed different values of the selected similarity exponent in either case (figure \ref{fig:varying_behavior_powerlaws}b). This behavior is in direct analogy with planar Marangoni filling flows, where \cite{temprano2024self} showed that $\beta=2$ for the quadratic case ($\theta=2$) and $\beta=3/2$ for flatter profiles ($\theta>2$), regardless of the specific value of $\theta$. However, in contrast with the planar case, here we also measured a different value of the exponent $\beta$ for every value of $\theta$, i.e., for every initial concentration profiles of a given polynomial order near the origin (figure \ref{fig:varying_behavior_powerlaws}c, d), suggesting the existence of a countably infinite number of similarity exponents. This is shown more clearly in figure \ref{fig:beta-vs-theta} by explicitly considering initial concentrations which scale as $\Gamma_0 \sim kr^\theta \text{ as } r\to{0}$. A monotonic dependence of $\beta$ on $\theta$ is observed, with $\beta$ quickly leveling off for increasingly flat initial conditions towards $\beta = 1.264 \pm 0.007$ as $\theta \to{\infty}$. 
\subsection{Discussion of numerical results} \label{subsec:discussion-filling}
Unlike self-similar solutions of the first kind, similarity solutions of the second kind are characterized by the lasting influence of features of initial conditions in the self-similar asymptotics \citep{barenblatt1996scaling}. Such problems may admit a spectrum of similarity exponents, which may be discrete or continuous \citep{barenblatt_self-similar_1972}. When the eigenspectrum of \eqref{eq:similarity-ODE} contains multiple admissible values of $\beta$, it is often shown by arguments of dynamic stability that only a small subset of values of $\beta$ is of physical relevance and is ultimately selected by the governing equations \eqref{eq:pde-with-closure} \citep{Eggers_2009}. 

For instance, for planar Marangoni filling flows, while a countably infinite number of exponents exists, stability arguments select only two physically relevant values of $\beta$ depending on the near-origin asymptotic behavior of the initial condition \citep{temprano2024self}. As previously mentioned, if the initial concentration profile behaves quadratically near the origin, the exponent $\beta=2$ prevails in the similarity solution. Conversely, for initial conditions which are flatter near the origin, the exponent $\beta=3/2$ is selected. 

It is interesting to note from figures \ref{fig:varying_behavior_powerlaws}a, b that a comparable selection criterion for $\beta$ exists for axisymmetric flows. That is, despite the non-locality of \eqref{eq:pde-with-closure}, the global features of the initial condition do not influence the singularity exponent as $t\to{t_*}$, only its local behavior seems to do so. However, unlike planar Marangoni filling flows, the dynamically selected value of $\beta$ is found numerically to admit a countably infinite number of stable values depending on the near-origin structure of the initial conditions (figure \ref{fig:beta-vs-theta}).

The variation of dynamically stable similarity exponents has been observed in other problems when the non-local character of the closure relation is changed \citep{eggers_selection_2020}. However, in our case, the nature of the transport equation and geometry remain the same across all trials, and the continuous similarity selection process is associated primarily with the local structure of the initial conditions. 

\section{Conclusion} \label{sec:conclusion}
Guided by recent progress on the dynamics of radially spreading surfactants on a deep, viscous subphase \citep{bickel2022exact}, we revisit axisymmetric Marangoni flows from the perspective of self-similarity. We consider both spreading and filling flows, the latter referring to flows in which a surfactant concentration distribution collapses to `fill' a region of high surface tension. Self-similar filling solutions have previously been identified for axisymmetric Marangoni flows on a thin subphase \citep{jensen_self-similar_1994} and for planar Marangoni flows on a deep subphase \citep{temprano2024self}; here we consider specifically the case of deep, axisymmetric filling flows.

In section \ref{sec:spreading-sol}, we revisit the axisymmetric Marangoni spreading problem of insoluble surfactants, and obtain a self-similar solution globally valid at long times $t\gg 1$ where the surfactant front is shown to spread as $\ell(t) \propto t^{1/3}$ (consistent with the dimensional predictions shown in table \ref{tab:surfactant-front-scaling-table}), using the formalism of dual integral equations \citep{sneddon1966mixed}. Solutions for this physical problem were previously identified by \citet{bickel2022exact} and more generally by \citet{caffarelli_nonlinear_2011, biler2011barenblatt}, using different methods.

In section \ref{sec:filling-sol}, filling flows of insoluble surfactant are considered, in direct analogy to the work in planar geometries of \citet{temprano2024self}. First investigated by \citet{jensen_self-similar_1994, thess1995viscous}, these flows have the feature of reaching singular structures (i.e., a cusp-like profile) in finite time, which are well described by self-similarity of the second kind \citep{Eggers_2009}. We show through direct numerical simulation of the governing equations \eqref{eq:pde-with-closure} that self-similar characteristics exist for such flows, and depend on the local structure of the initial concentration profile, consistent with the notion of self-similarity of the second kind \citep{barenblatt1996scaling} and in direct analogy to the filling solutions identified in planar domains \citep{temprano2024self}. In contrast to planar filling flows, where only two values of the similarity exponent are stable for filling flows, the numerical results point to an infinite number of stable exponents depending on the local behavior of the initial concentration profile, see figure \ref{fig:beta-vs-theta}.

\begin{bmhead}[Acknowledgments.]
We thank Dimitrios Fraggedakis and Jun Eshima for helpful discussions. We also thank Connor Roeser for assistance in the early development of the project. We acknowledge the use of Fabio Crameri's color maps in many of the figures in this manuscript \citep{crameri2018scientific}, as well as OpenAI's GPT-5.6 Sol model for assistance with figure layout and formatting. H.A.S. acknowledges support from Grant No. CBET-2246791 from the U.S. National Science Foundation.
\end{bmhead}

\begin{bmhead}[Declaration of interests.]
The authors report no conflict of interest.
\end{bmhead}

\begin{appen}

\section{Numerical evaluation of the closure relation}\label{apx:numerics_closure}

We first note that the closure given by \eqref{eq:closure-axisymmetric}, while nonlocal, is still linear. In addition, the operator $\mathcal{H}_1^{-1}\mathcal{H}_0$ applied to a constant function yields zero, i.e. a uniform concentration produces no velocity field $u_s(r,t)=0$. These two facts imply that we can assume, without loss of generality, a decaying concentration profile $\Gamma(r,t)$ as $r\to\infty$ for the purpose of evaluating the closure relation. Assuming also that $\Gamma(r,t)$ is well-behaved so that \eqref{eq:closure-axisymmetric} converges, we first rewrite it as the limit
\begin{equation}
    u_s(r,t) = \lim_{S\to\infty}\int_0^S\int_0^\infty sr' J_1(s r)   J_0(s r')\Gamma(r',t) \dd{r'}\dd{s}.
\end{equation}
Since Bessel functions are bounded (i.e. $|J_n|\le1$), the integrand of this expression is also bounded by $s\,r'|\Gamma(r',t)|$, so as long as $r'\Gamma(r',t)$ is integrable on $r'\in(0,\infty)$, one can use Fubini's theorem to exchange the order of integration and arrive at 
\begin{equation}
    u_s(r,t) = \lim_{S\to\infty}\int_0^\infty r'\,\Gamma(r',t)\,
    \left(\int_0^S s\,J_1(sr)\,J_0(sr')\dd{s}\right)\dd{r'}.
\end{equation}
The kernel in parentheses has no pointwise limit as $S\to\infty$, since its integrand oscillates without decaying in $s$, but it does converge in the sense of distributions in $r'$. After introducing the substitution $\kappa=sr'$, we obtain
\begin{equation}
    u_s(r,t) = \lim_{S\to\infty}\int_0^\infty \,\frac{\Gamma(r',t)}{r'}\,
    \left(\int_0^{Sr'} \kappa\, J_1\!\left(\frac{r}{r'}\,\kappa\right) J_0(\kappa)\dd{\kappa}\right)\dd{r'},
\end{equation}
for which the inner integral in the limit $S\to\infty$ has been evaluated in the sense of distributions by \cite{Kellendonk2009-xj}. Following their results, and expressing hypergeometric functions in terms of complete elliptic integrals \citep{NIST:DLMF}, we can write the outer integral as a principal value integral with a singular kernel
\begin{equation} \label{eq:closure}
    u_s(r,t) = \dashint{0}{\infty} \Gamma(r',t) \left\{\frac{r'}{\upi{r}}\left[\frac{\KK{\frac{4rr'}{(r+r')^2}}}{r+r'}+\frac{\EE{\frac{4rr'}{(r+r')^2}}}{r-r'}\right]\right\} \dd{r'}, 
\end{equation}
with 
\begin{equation}
    \KK{m} = \int_0^{\upi/2}\frac{\dd{\theta}}{\sqrt{1-m\sin^2\theta}}\,, \qquad
    \EE{m} = \int_0^{\upi/2}\sqrt{1-m\sin^2\theta}\,\dd{\theta},
\end{equation}
the complete elliptic integrals of the first and second kind, respectively. We note that the kernel in \eqref{eq:closure}, up to a constant prefactor, has been previously obtained in other problems unrelated to surfactants but also involving creeping flows in semi-infinite domains \citep[eq.~C10b]{Lister1989-so}.

Equation \eqref{eq:closure}, which only involves a single integral, is the expression we use to numerically evaluate the interfacial velocity $u_s(r,t)$ in terms of the concentration $\Gamma(r,t)$. The invariance of \eqref{eq:closure} with respect to spatial rescalings, which becomes evident noting that $r\to{L}{r}$ and $r'\to{L}{r'}$ (with $L>0$ an arbitrary constant) leave the expression unchanged, motivates a change of variables
\begin{equation}
    \theta \defeq \log(r/L), \qquad \theta' = \log(r'/L),
\end{equation}
which leads to 
\begin{align}
    u_s(\theta,t) &= \dashint{-\infty}{\infty} \Gamma(\theta',t) \left\{\dfrac{e^{-(\theta-\theta')}}{\upi}\left[\dfrac{\KK{\sech^2\left(\frac{\theta-\theta'}{2}\right)}}{e^{\theta-\theta'}+1}+\dfrac{\EE{\sech^2\left(\frac{\theta-\theta'}{2}\right)}}{e^{\theta-\theta'}-1}\right]\right\}\dd{\theta'}.
\end{align}
Note that the spatial domain is now the whole real line $\theta\in(-\infty,\infty)$, and that the expression is in convolution form, with the kernel now depending on $\theta-\theta'$ rather than on $\theta$ and $\theta'$ separately. This structure is key for the fast evaluation of the closure in Fourier space using the convolution theorem.

However, naive evaluation in Fourier space presents two obstacles. First, while the profile $\Gamma(\theta,t)$ can be taken to decay as $\theta\to\infty$ (i.e. $r\to\infty$), it does not in general decay as $\theta\to-\infty$ (i.e. $r=0$), and both of these conditions are required to evaluate its Fourier transform. To address this issue, we split the concentration as
\begin{align}
      \Gamma(r,t) &\defeq \Gamma(0)e^{-(r/\ell)^2} + \tilde{\Gamma}(r,t),
\end{align}
where $\ell$ is a positive constant that is taken as the length of a few grid cells in the numerical simulation. Thanks to the linearity of the closure, this decomposition allows us to write
\begin{align}
    u_s(r,t) &= \Gamma(0)\,\mathcal{H}_1^{-1}\mathcal{H}_0\left[e^{-(r/\ell)^2}\right] + \tilde{u}_s(r,t),
\end{align}
where the first term can be computed exactly as
\begin{equation}
    \mathcal{H}_1^{-1}\mathcal{H}_0\left[e^{-(r/\ell)^2}\right] = \dfrac{\sqrt{\pi}r}{2\ell}e^{-(r/\ell)^2}\left[I_0\left(\dfrac{r^2}{2\ell^2}\right)-I_1\left(\dfrac{r^2}{2\ell^2}\right)\right]
\end{equation}
using known Hankel transform pairs \citep{NIST:DLMF}, and where $I_0(\,\cdot\,)$ and $I_1(\,\cdot\,)$ are modified Bessel functions. This reduces the problem into computing $\tilde{u}_s(r,t)=\mathcal{H}_1^{-1}\mathcal{H}_0\left[\tilde{\Gamma}(\theta,t)\right]$, where $\tilde{\Gamma}(\theta,t)$ now decays at both ends of the far field $\theta\to\pm\infty$. 

The second numerical obstacle is evident from the asymptotic expansions of the kernel of \eqref{eq:closure}, 
\begin{align}\label{eq:kernel}
    \mathcal{K}(\theta)\defeq \dfrac{e^{-\theta}}{\upi}\left[\dfrac{\KK{\sech^2\left(\frac{\theta}{2}\right)}}{e^{\theta}+1}+\dfrac{\EE{\sech^2\left(\frac{\theta}{2}\right)}}{e^{\theta}-1}\right] = 
    \begin{cases}
    -\dfrac{1}{2}e^\theta + O(e^{2\theta}) &\text{as }\theta\to-\infty,\\[10pt]
    \dfrac{1}{\pi\theta} - \dfrac{\log|\theta|}{2\pi} + O(1) &\text{as }\theta\to{0},\\[10pt]
    e^{-2\theta} + O(e^{-4\theta}) &\text{as }\theta\to\infty
    \end{cases}
\end{align}
which show that, while the kernel decays at both ends $\theta\to\pm\infty$, it also has two singular terms as $\theta\to{0}$. To address this issue, we split the kernel as
\begin{align}
    \mathcal{K}(\theta) &\defeq \mathcal{K}_R(\theta) + \mathcal{K}_{S1}(\theta) + \mathcal{K}_{S2}(\theta),
\end{align}
where 
\begin{subequations}
\begin{align}
    \mathcal{K}_{S1}(\theta) &\defeq \dfrac{e^{-(\theta/\sigma)^2}}{\pi\theta},\\
    \mathcal{K}_{S2}(\theta) &\defeq \dfrac{E_1\left((\theta/\sigma)^2\right)}{4\pi},
\end{align}
\end{subequations}
are chosen to have the same singular behavior as $\theta\to{0}$ as each singular term in \eqref{eq:kernel}, respectively, while decaying quickly in the far field to leave the far-field behavior of $\mathcal{K}_R(\theta)$ unaffected. Here, $E_1(\,\cdot\,)$ is the exponential integral, $E_1(x)\defeq\int_x^\infty t^{-1}e^{-t}\dd{t}$, which behaves as
$E_1(x) = - \log x -\gamma_E + O(x)$ as $x\to0$, with $\gamma_E$ the
Euler--Mascheroni constant, and $\sigma$ is a positive constant taken as the length of a few grid cells (in the transformed $\theta$ coordinate) in the simulations. 

With this decomposition, we can write $\tilde{u}_s(\theta,t)$ as a sum of convolutions
\begin{equation}
    \tilde{u}_s = \tilde{\Gamma}\ast\mathcal{K}_R + \tilde{\Gamma}\ast\mathcal{K}_{S1} + \tilde{\Gamma}\ast\mathcal{K}_{S2}, 
\end{equation}
which, after taking the Fourier transform $\mathcal{F}$ and applying the convolution theorem, can be expressed as
\begin{equation}
    \mathcal{F}[\tilde{u}_s] = \mathcal{F}[\tilde{\Gamma}]\cdot\mathcal{F}[\mathcal{K}_R] + \mathcal{F}[\tilde{\Gamma}]\cdot\mathcal{F}[\mathcal{K}_{S1}] + \mathcal{F}[\tilde{\Gamma}]\cdot\mathcal{F}[\mathcal{K}_{S2}]. 
\end{equation}
The computation of the Fourier transforms of the singular functions $\mathcal{K}_{S1}$ and $\mathcal{K}_{S2}$ can be avoided by noting that they can be calculated exactly as
\begin{subequations}
\begin{align}
    \mathcal{F}[\mathcal{K}_{S1}](\omega) &= -\mathrm{i}\,\mathrm{erf}\left(\dfrac{\sigma\omega}{2}\right),\\
    \mathcal{F}[\mathcal{K}_{S2}](\omega) &= \dfrac{1}{2|\omega|}\mathrm{erf}\left(\dfrac{\sigma|\omega|}{2}\right).
\end{align}
\end{subequations}
Consequently, we finally obtain
\begin{equation} \label{eq:num_fourier}
    \tilde{u}_s = \mathcal{F}^{-1}\left[\mathcal{F}[\tilde{\Gamma}]\cdot\left\{\mathcal{F}[\mathcal{K}_R] -\mathrm{i}\,\mathrm{erf}\left(\dfrac{\sigma\omega}{2}\right) + \dfrac{1}{2|\omega|}\mathrm{erf}\left(\dfrac{\sigma|\omega|}{2}\right)\right\}\right]. 
\end{equation}
The only numerical computations are therefore a Fourier transform of $\mathcal{K}_R$ (which can be precomputed and use throughout the simulation since $\mathcal{K}_R$ does not change in time), a Fourier transform of $\tilde{\Gamma}$, and an inverse Fourier transform of the expression in square brackets in \eqref{eq:num_fourier}. These operations involve well-behaved, decaying functions, and efficient numerical computation with standard Fast Fourier Transform (FFT) libraries is straightforward.

\section{Extrapolation to grid-independent results}\label{apx:numerics_extrapolation}
In \S \ref{subsec:num-analysis-results}, properties of the singular structures corresponding to different initial concentration profiles $\Gamma_0$ are reported, including the similarity exponent $\beta$ inferred from each singularity (see table \ref{tab:pre-and-post-singularity-beta} or figure \ref{fig:beta-vs-theta}, for instance) and the time at which such a singularity occurs. Here, we outline how such estimates were obtained from individual simulations involving different grid-refinement parameters. As explained in \S \ref{subsec:num-analysis-methods}, a finite-volume scheme with a one-dimensional radial grid is implemented, with the first cell face defined at $r=0$, and the last cell center defined at a large radius $L$, which is kept constant across all simulations. All cell centers are placed such that $r_{n+1} / r_n = \lambda = \text{const}$. The position of the first cell center, $r_{\min}$, and the number of cells in the grid $n_x$ are directly varied across different simulations. The corresponding ratio $\lambda$ is then determined as
\begin{equation}
    \lambda = \left(\frac{L}{r_{\min}}\right)^{1/(n_x - 1)}.
\end{equation}
Effectively, $r_{\min}$ and $\lambda$ are the only parameters that control the refinement of the mesh used for each numerical trial. The continuum limit then corresponds to $r_{\min} \to{0}$ and $\log \lambda \to{0} $ \citep{hinch2020numerical}. Figure \ref{fig:extrapolation-example} displays the typical grid convergence of $t_*, \beta$ for a representative initial condition. Note that each data point shown is obtained using one of the measurement methods illustrated in figure \ref{fig:selection-criterion}. The process of extrapolating numerical data to grid-independent results is two-fold: first a linear extrapolation in $\log\lambda$ for each fixed value of $r_{\min}$ is performed to $\log \lambda = 0$, after which these intercepts are extrapolated in $r_{\min}$ to $r_{\min} = 0$. This yields a single grid-independent result, labeled $\alpha^\infty$ in the insets of figure \ref{fig:extrapolation-example} for any measured quantity $\alpha$. 
\begin{figure}
  \centerline{\includegraphics[width=1.0\textwidth]{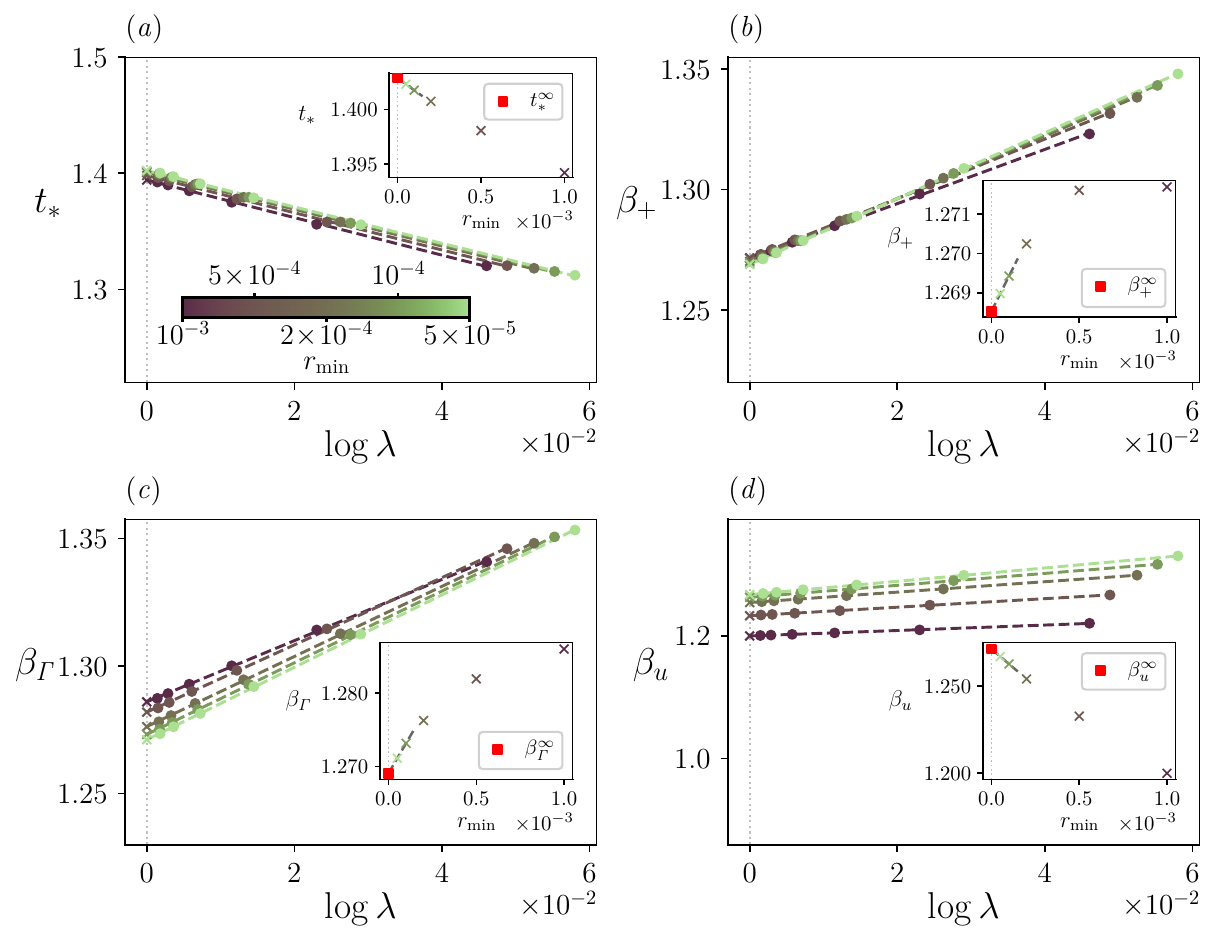}}
  \caption{Grid-independent values of relevant parameters for $\Gamma_0 = r^6 / (1+r^6)$. Each color corresponds to simulations performed at a fixed value of $r_{\min}$, while the points in each color series were obtained using a different value of $n_x$, and hence $\lambda$. Each inset shows the extrapolation in $r_{\min}$ of the intercepts at $\log\lambda = 0$ shown in the parent plot. Panels (\textit{a}), (\textit{c}), and (\textit{d}) are constructed from the `closure' solution of $\Gamma$, and pertain to the measurement method presented in figure \ref{fig:selection-criterion}a. Panel (\textit{b}) is constructed from the `leveling' behavior of $\Gamma$, and pertains to the measurement method presented in figure \ref{fig:selection-criterion}b.}
  \label{fig:extrapolation-example}
\end{figure}
As mentioned in \S \ref{subsec:num-analysis-methods}, the values of $\beta$ and $t_*$ corresponding to each measurement method in figure \ref{fig:selection-criterion} are inferred by applying a power-law fit over a given region of interest. While the location of the fitting window is chosen systematically by \eqref{eq:pre-singularity-selection-criterion}, its size is prescribed. To mitigate such subjectivity, the process outlined above is repeated for several fitting window sizes, and grid-independent results are obtained for each window size. The final results presented in table \ref{tab:pre-and-post-singularity-beta} and figure \ref{fig:beta-vs-theta} are simply the average of those results, and the uncertainty reported is the largest absolute difference between the average and any single grid-independent value.

\end{appen}\clearpage

\bibliographystyle{jfm}
\bibliography{jfm}


\end{document}